# Power Law Distribution of the Frequency of Demises of U.S Firms[1]

William Cook and Paul Ormerod

Volterra Consulting, 121 Mortlake High Street, London SW14 8SN, UK

e-mail:     wcook@volterra.co.uk, pormerod@volterra.co.uk

April 2002

*Abstract*

*Both theoretical and applied economics have a great deal to say about many aspects of the firm, but the literature on the extinctions, or demises, of firms is very sparse. We use a publicly available data base covering some 6 million firms in the US and show that the underlying statistical distribution which characterises the frequency of firm demises - the disappearances of firms as autonomous entities - is closely approximated by a power law. The exponent of the power law is, intriguingly, close to that reported in the literature on the extinction of biological species.*

The purpose of this paper is to provide empirical evidence on the statistical distribution which appears to characterise the demises of US firms, across the entire universe of such firms using a publicly available data base. The evidence is consistent with the hypothesis that the data follow a power law distribution. The observation of power-law distributions (fractal behaviour) in a system's macroscopically observable quantities is a characteristic property of many-body systems representing the effects of complex interactions amongst the constituents of the system. Recent evidence that related aspects of economic activity are consistent with power distributions at the aggregate level is given, for example, by [1] on the growth of firms, [2] on the sizes of firms and [3] on the duration of recessions in the Western market economies.

Economic theory has a great deal to say about many aspects of the firm. Empirical studies of firm growth and size have emphasised the importance of stochastic influences, from

---

[1] we are grateful to the Institute for Complex Additive Systems Analysis of the New Mexico Institute of

the initial work of Gibrat [4] through classic papers in the 1950s and 1960s [for example 5,6], to more recent contributions such as [1, 2]. However, the literature on demises of firms is surprisingly sparse.

The disappearance of a firm as an autonomous entity - its demise - can take place for a variety of reasons, such as merger, take over and bankruptcy. The proximate reasons for demises over the 1912-95 period amongst the 100 largest industrial companies in the world in 1912 are given in [7], and similar evidence over the 1919-79 period of the 100 largest companies in the US on a decade by decade basis is provided by [8]. Very small firms constitute the overwhelming majority of incorporated businesses in terms of numbers, many of which are controlled by a single shareholder. Here, the reasons for a firm ceasing to exist can be even more diverse. In addition to bankruptcy, the owner may, for example, close a firm down simply because he or she has decided to focus attention on a different area of business. A small number of empirical studies which relate firm demises within individual industries to factors such as the number of firms in an industry at the time the firm enters are cited in [9], but no general analysis of demises appears to be available.

The data we examine comes from the American Office of Advocacy. Recorded in the data are the frequency of firm 'deaths' on an annual basis from 1989 through 1997 split into nine different industrial sectors for each of the 51 states[2]. This gives rise to 459 series of 9 annual observations in 51 states, a total of 4131 observations in total. We describe each of these observations as a 'group'. In other words, each group specifies a particular industry in a particular year in a particular state.

Data is also provided on the total number of firms for each of the 4131 groups. The number varies enormously, with the smallest being mining in Hawaii in 1997, with just six firms. The largest is service sector firms in California, with 265,710, also in 1997. To avoid any potential problems which might arise from the very small size of some

---

Mining and Technology for financial support towards this research
[2] District of Columbia is included as a state in this data set

groups, we exclude from the analysis any group which has in total less than 200 firms. This leaves out 221 observations, just over 5 per cent of the total, leaving a sample of 3910 observations to be analysed.

The average total number of firms across the United States over this period was 5.73 million, of which on average 611,000 died each year. The total number of firm demises varied very little from year to year, the maximum being 664,000 in 1996 and the minimum 577,000 in 1994. The number of demises bears little connection with the overall state of the economy. In the recession year 1991, for example, real GDP fell by 0.5 per cent and in the boom year 1997 it grew by 4.3 per cent. Yet the number of demises was very similar in both years, being 630,000 and 648,000 respectively.

The natural focus of analysis across the different groups is not, however, the total number of demises in each group but demises as a proportion of the total number of firms in each group. This eliminates the expected linearity between number of demises and group size, the correlation between the proportion of firms disappearing and the size of the group being only -0.02. Figure 1 plots the histogram of 'death' proportions.

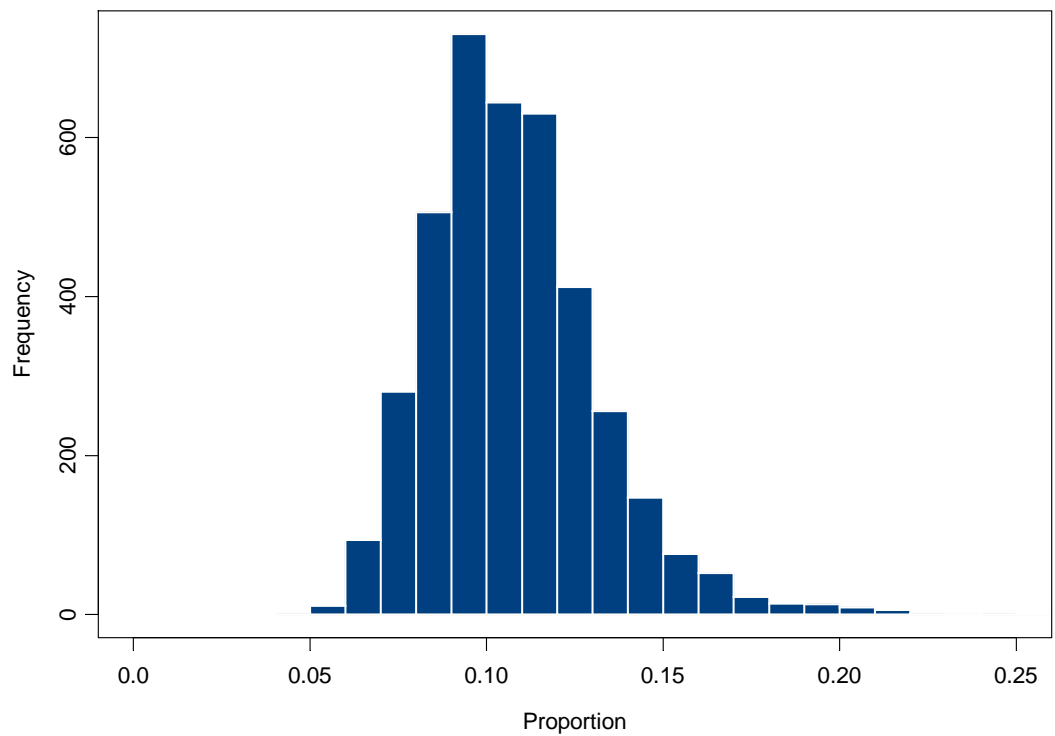

**Figure 1:** *Histogram of firm demises as a proportion of the total number of firms in the group. The group is defined by year, industry and state. Groups with less than 200 firms in total are omitted.*

At first sight, the data appears to be approximated closely by the lognormal distribution. The null hypothesis of lognormality is only rejected at p = 0.0009 using the Kolmogorov-Smirnov test.

However, this could be distinctly misleading. Each observation on the number of demises relates to the total number over the course of a year in any particular group. In other words, it is a *temporal* aggregation across a year of the demises which take place in any given category on each working day, approximately 250 per year.

The distribution we observe from the annual data is not necessarily that which gives rise to the data as it is actually generated on a daily basis. If day-to-day disappearances of firms are independent and identically distributed, we can treat each annual observation,

$A_i$, as being the sum of N independent random variables. By the Central Limit Theorem, sums of iid variables of finite variance will converge towards the Gaussian distribution as N increases. But the actually observed annual data is very definitely non-Gaussian (the null hypothesis of normality is rejected on a Kolmogorov-Smirnov test at p = 0.0000).

Indirect evidence on the independence of the daily events whose sum makes up the annual observations is available from the time-dependency of the annual data. The correlation between the current year's proportion of demises and the previous year's is 0.530 for the aggregate data - for the total number of demises divided by the total number of firms. However, this falls to an average of 0.265 across the 9 industries across time, and to an average of 0.161 across the 451 separate time-series for each industry in each state. In other words, the correlation over time between observations falls as the level of disaggregation of the data increases. This implies that the assumption that the daily observations - when the data is temporally disaggregated - of firm demises are independent is not an unreasonable one to make.

The assumption that the daily observations are identically distributed need not hold exactly for the Central Limit Theorem to be applicable. Rather, the variance of any separate distributions which exist must have variances which are not too dissimilar, so that no single variance dominates over all the others [10]. Given that the proportions of demises is in [0,1], again this does not seem an unreasonable assumption.

Only when the data is restricted to a sub-set of only 221 observations immediately around the mean, just over 5 per cent of the total, is normality not rejected at the conventional level of p = 0.05. In other words, the actual distribution of firm 'deaths' which we observe with data aggregated over a year converges only very slowly to a Gaussian. Each observation consists of the sum of some 250 daily events, yet despite this only a small fraction of the data is well described by the Gaussian distribution. The sum of independently, identically distributed random variables is known to converge only very slowly to a Gaussian when the underlying distribution follows a power law[3] [10]. We therefore postulate that the underlying distribution follows a truncated power law of the form:

$$P(X = i) = \frac{(i+1)^{-r}}{\sum_{j=0}^{m}(j+1)^{-r}} \qquad i = 0, \ldots, m \qquad (1)$$

In this form, the distribution is dependent on two parameters, r and m. The rate at which the distribution falls is described by the parameter r, tailing off faster as r increases. The maximum value the distribution can take is given by m. We introduce m because allowing limitless demises per time period does not model a possible real life scenario. The generated data was scaled to the range of proportions observed.

We experimented with a range of values for m and r (for r > 2). The maximum annual number of demises observed in any group in the data was 30,366, a daily average rate of some 121, though for 99 per cent of the groups the implied daily average is less than 50. We examined the outcome with m chosen over the range 10 to 500. In each case, we drew 250 values, and summed them, and repeated the procedure 3,910 times, to obtain a series the same length as the number of observations in the actual data set examined. The best fit to the actual data was obtained with m = 230 and r = 2.1. The correlation between the (sorted) artificially-generated data set and the actual data is 0.995, and the null hypothesis that the two distributions are the same is only rejected on a Kolmogorov-Smirnov test at p = 0.115.

In terms of robustness of the results, the null hypothesis that the two distributions are the same is not rejected on a Kolmogorov-Smirnov test at the standard level of significance of p = 0.05 for values of m between 120 and 310. Interestingly, the literature on the extinction rates of biological species reports that the frequency/size relationship can also

---

[3] when the variance is finite. With a power law with infinite variance, the convergence is to the more general class of Levy distributions, of which the Gaussian is a special case

be approximated by a power law with an exponent close to 2 in absolute value [for example, 11].

The two sorted sets of data are plotted in Figure 2.

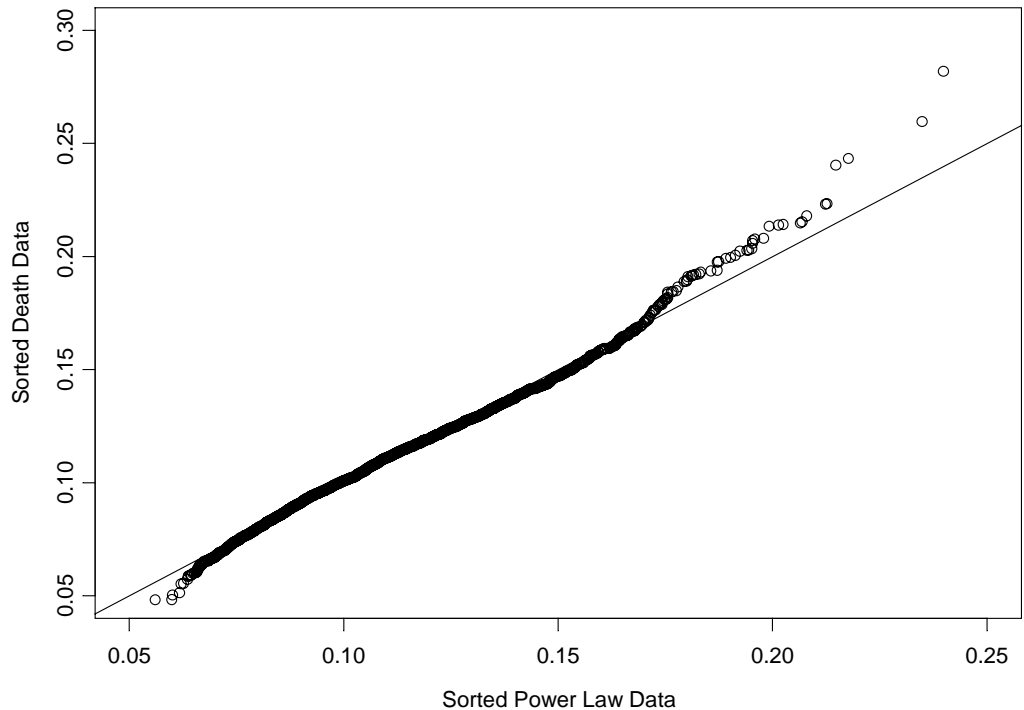

**Figure 2**   *Data generated from the truncated power law (1) with m = 230 and r = 2.1 plotted against the actual data, both data sets being sorted by size.  Each observation of the generated data is the sum of 250 observations drawn from (1).*

We notice small deviations from the true distribution in the lower and upper tails.  It is likely that the swing at the lower end is due to the small group size association still remaining.

In summary, the underlying statistical distribution which characterises the frequency of firm demises -the disappearance of a firm as an autonomous entity- in the United States is approximated well by a power law. The exponent of the power law is, intriguingly, close to that reported in the literature on the extinction of biological species